\documentclass[draft]{agujournal2019}
\usepackage{url} 
\usepackage{lineno}
\usepackage{soul}
\usepackage{amsmath} 
\usepackage{booktabs}
\usepackage{multirow}
\linenumbers
\nolinenumbers

\journalname{}

\begin{document}

\title{Physics-Informed Deep Operator Learning for Computational Hydraulics Modeling}

\authors{Xiaofeng Liu\affil{1,2} and Yong G. Lai\affil{3}}

\affiliation{1}{Department of Civil and Environmental Engineering, The Pennsylvania State University, University Park, PA, USA}
\affiliation{2}{Institute of Computational and Data Sciences, The Pennsylvania State University, University Park, PA, USA}
\affiliation{3}{Technical Service Center, U.S. Bureau of Reclamation, Denver, CO, USA}

\correspondingauthor{Xiaofeng Liu}{xzl123@psu.edu}

\begin{keypoints}
\item Physics-informed DeepONet framework is developed for computational hydraulics modeling based on the 2D shallow water equations
\item Physics constraint comes with the cost of higher errors for in-distribution cases because of the inconsistency between discrete and continuous governing equations
\item Physics constraint improves the generalization capability of the model for out-of-distribution cases
\end{keypoints}

\begin{abstract}
Traditional 2D hydraulic models face significant computational challenges that limit their applications that are time-sensitive or require many model evaluations. This study presents a physics-informed Deep Operator Network (DeepONet) framework for computational hydraulics modeling that learns the solution operator of the 2D shallow water equations (SWEs) to create fast surrogate models. The framework can operate in two modes: a purely data-driven SWE-DeepONet that learns from numerical solver such as SRH-2D, and a physics-informed PI-SWE-DeepONet that additionally incorporates the continuous SWEs as constraints during training. Based on a real-world case, steady flows in a reach of the Sacramento River in California, it is demonstrated that PI-SWE-DeepONet possesses much enhanced prediction capability than SWE-DeepONet when applied to out-of-distribution scenarios. The physics-informed model is shown to exhibit slower error growth and larger breakdown distances in comparison with SWE-DeepONet. The gain of the physics-informed training, however, comes with costs, chief among which are the simulated results have slightly higher errors for in-distribution cases. It reflects the existence of a tension between the two competing training objectives: fitting the results from the traditional hydraulic model and satisfying the continuous governing equations. In this study, guidelines are developed for selecting the appropriate approach based on a real-world case: PI-SWE-DeepONet is preferred for out-of-distribution predictions, uncertain training data, or when physical consistency is a priority, while SWE-DeepONet is recommended if the modeling objective is to replicate faithfully the traditional hydraulic model results within the training distribution. Other challenges are also discussed, such as the loss weighting approach.
\end{abstract}

\section{Introduction}\label{sec:introduction}

Two-dimensional (2D) depth-averaged hydraulic models solve the shallow water equations (SWEs), which are derived from depth-averaging the Navier-Stokes equations under the hydrostatic pressure assumption. They have been widely used in recently years due to the availability of models such as SRH-2D (Sedimentation and River Hydraulics 2D) \cite{lai2010two} and HEC-RAS 2D (Hydrologic Engineering Center's River Analysis System) \cite{brunner2016hec}. They have applied to simulate complex flows along river corridors and in estuaries and coastal regions. The ability to accurately predict water surface elevations, flow velocities, and inundation extents under varying boundary conditions and forcing scenarios is essential for emergency response planning, infrastructure design, river restoration, and water resources management.

Despite their widespread adoption and proven reliability, traditional 2D hydraulic models face computational challenges that have placed a limit on their applicability range. High-fidelity simulations of large-scale flood events or complex river systems can require hours to days of computational time, even on modern high-performance computing systems. This computational burden becomes a bottleneck in applications requiring many model runs for evaluation, such as uncertainty quantification through Monte Carlo simulations, real-time flood forecasting, parameter calibration and optimization, design alternative evaluation, and risk assessment. In general, the computer runtime scales with domain extent, mesh resolution, simulation duration, and the complexity of the physical processes. For time-sensitive applications such as flood forecasting, the prolonged runtime is unacceptable and has prevented the use of high-resolution 2D models, resulted in the use of faster but less reliable 1D models or simplified approaches.

To address the computational bottleneck, high performance computers using massively parallel computational processing units (CPU) and graphical processing units (GPU) have been used \cite{YongEtAl2025EWRI}. The gain, however, is insufficient despite improvements. In recent years, researchers and practitioners have increasingly turned to surrogate modeling approaches that can approximate the behavior of traditional hydraulic models at a fraction of the computational cost. Surrogate models learn the input-output relationship of the original model from a limited set of training simulations, then provide rapid predictions for new inputs without having to solve the full partial differential equation (PDE) system. It is widely accepted that the most promising surrogate models are based on deep machine learning techniques. Such models have been explored using the convolutional neural network (CNN) for spatial variation, long short-term memory (LSTM) network for temporal evolution, and fully connected neural network for parameter-to-output mapping \cite{songSurrogateModelShallow2023a,haces-garciaRapid2DHydrodynamic2025}. However, these traditional machine learning approaches typically require re-training for each new set of initial or boundary conditions, and therefore lack theoretical guarantees about their predictive capabilities. 

A fundamentally different but powerful paradigm for surrogate modeling of PDE-based systems has emerged in recent years - the field of operator learning which learns mappings between infinite-dimensional function spaces rather than finite-dimensional vectors. Operator learning represents a transformative approach in scientific machine learning (SciML) that addresses the fundamental challenge of learning solution operators for parametric partial differential equations \cite{lu2021deep,kovachki2024operator,li2021fourier}. Major operator learning architectures include DeepONet \cite{lu2021deep,lu2022comprehensive}, Fourier Neural Operator (FNO) \cite{li2021fourier,li2021neural}, Physics-Informed Neural Operators (PINO) \cite{li2021physics,wang2021learning,goswami2022physics}, Graph Neural Operators (GNO) \cite{li2023geometry}, and U-Net Enhanced Operator Networks (e.g., U-DeepONet, U-FNO) \cite{rahman2023u,wen2022u}. Operator learning enable the surrogate to generalize across different initial conditions, boundary conditions, and physical parameters without retraining. Given a PDE with varying inputs, the solution operator maps these input functions to the corresponding solution functions. For SWEs, this means learning an operator that can predict the water depth and velocity fields for any given set of  boundary conditions, initial conditions, or bathymetry - a significant predictive capability needed to overcome the traditional methods.

The Deep Operator Network (DeepONet), introduced in \citeA{lu2021deep}, was among the first to demonstrate the feasibility of learning solution operators for PDEs \cite{lu2022comprehensive}. DeepONet employs a two-branch architecture inspired by the universal approximation theorem for operators: a branch network processes discretized input functions (e.g., boundary conditions in the form of inflow hydrographs), while a trunk network takes spatiotemporal coordinates as input. The output is computed through an inner product of the branch and trunk network outputs, allowing predictions at any query locations. This architecture provides several advantages: flexibility in handling arbitrary input functions, capability for prediction at any query location in both space and time, and strong theoretical foundation, based on universal approximation theory, for operators \cite{lu2022comprehensive,kovachki2024operator}. DeepONet has demonstrated effectiveness across diverse applications, especially for problems where the input-output relationship exhibits smooth, learnable patterns \cite{lu2022comprehensive}.

A further development is the physics-informed (PI) DeepONet architecture which extends the data-driven framework by incorporating governing PDEs directly into training through a loss term that enforces the equations at collocation points \cite{wang2021learning,goswami2022physics}. The physics-informed training process allows the model to learn from both available simulation data and the underlying flow physics, creating a more robust and reliable surrogate for applications. This approach has been shown to reduce the need for a large set of training data and potential to improve the predictive capability to untrained scenarios \cite{wang2021learning,li2021neural,boule2024mathematical,chen2024}. Recent applications to computational hydraulics include shallow water dynamics forecasting, streamflow prediction, and groundwater flow modeling \cite{sun2024bridging,peherstorfer2024learning,rivera2025neural,keppler2025}.

In this study, the physics informed DeepONet framework is adopted so that the SWEs of the traditional 2D hydraulic models are incorporated as constraints during data training. By embedding the SWEs directly into the network architecture and training procedure, we aim to enhance the predictive capability of the surrogate model so that it is valid for out-of-distribution cases (those not in the training data distribution) despite the use of a limited set of training data. Further, we aim to explore whether the expected advantages of the physics informed method may be realized in practical applications – a section of the Sacramento River in California is selected as a demonstration case. When real-world cases are the application target with physics-informed operator learning, several theoretical and practical challenges start to appear and need to be addressed. The questions addressed in this study include:
\begin{itemize}
    \item How can the physics-informed DeepONet framework be designed to facilitate practical adoption in real-world engineering applications?
    \item How to properly balance the data loss and the physics loss? And within SWEs, how to  balance the losses (residuals) of the continuity equation, $x$-momentum equation, and $y$-momentum equation?
    \item Is physics constraint always good and necessary?
\end{itemize}

Technical and implementation challenges exist for real-world applications. Often, literature on operator learning focuses on simple canonical problems and conclusions from these simple cases may not hold with real-world applications. At present, there is a big gap between the academic benchmarks and the real-world cases. Modern-day practical hydraulic models mostly operate on unstructured meshes that conform to complex natural geometries due to irregular river planform, widely varying floodplains with vegetations, and instream structures. Some operator learning architectures such as FNO require regular, structured grids, making them unsuitable for the irregular domains commonly used in hydraulic engineering \cite{li2023geometry,lu2022comprehensive}. Real-world applications also involve complex boundary conditions, including multiple inlet discharges, stage-discharge relationships at outlets, and spatially varying bathymetry and roughness fields that must be properly incorporated into the operator learning framework. 

In this study, the above challenges are addressed by developing a flexible framework and workflow. An automated process is developed for all four stages of the modeling process: generating the training data, designing the operator learning architecture, training the model, and testing the performance. In the process, guidelines may be developed in real-world applications. A key outcome of the present study is the development of a tool that can be readily applied by engineers to real-world problems, enabling rapid scenario analysis, uncertainty quantification, risk analyses, and real-time forecasting, while maintaining the accuracy and physical consistency required for engineering decision-making.

The rest of the paper is organized as follows. The physics-informed DeepONet framework is first introduced. The application example is presented. A discussion on the results and the limitations of the framework is then provided. 

\section{Methodology}\label{sec:methodology}


\subsection{Physics-Informed DeepONet Framework}\label{subsec:physics-informed-DeepONet-framework}

\begin{figure}[htp]
    \centering
    \includegraphics[width=0.9\textwidth]{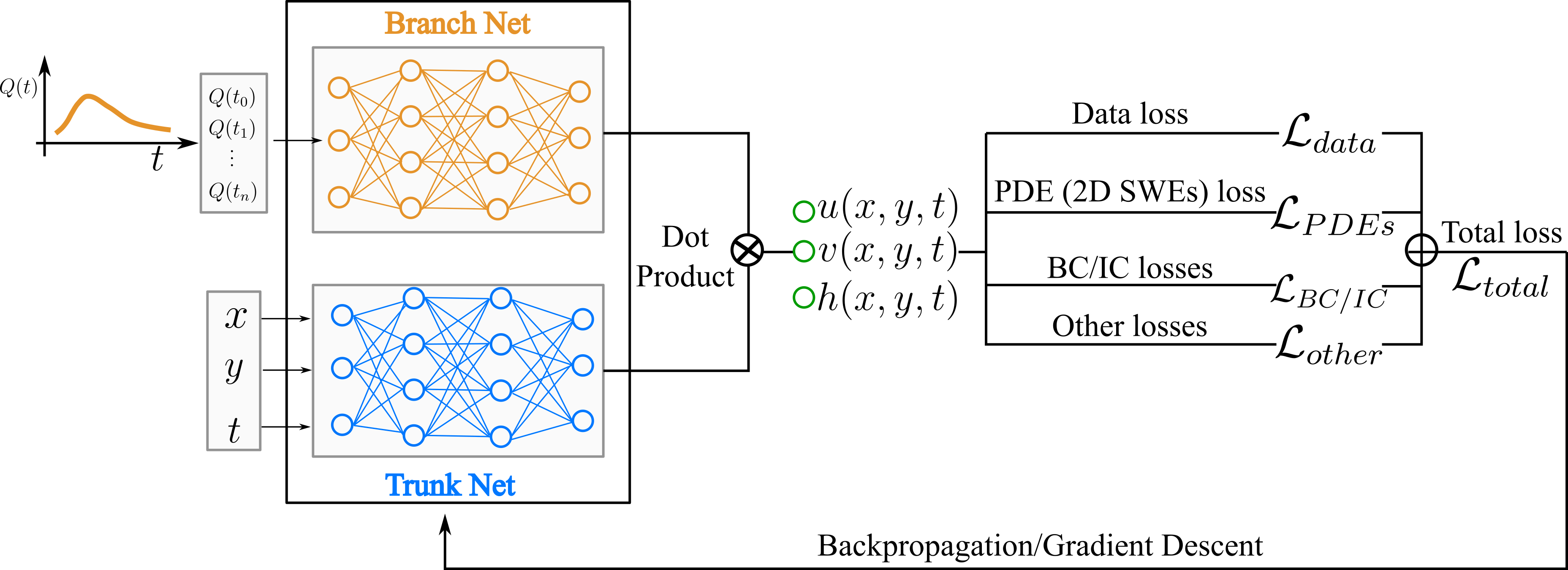}
    \caption{The physics-informed DeepONet framework for open channel flows governed by the 2D shallow water equations.}
    \label{fig:physics-informed-DeepONet-framework}
\end{figure}

The DeepONet framework employs a two-branch neural network architecture designed to learn the solution operator of the 2D SWEs. As illustrated in Figure~\ref{fig:physics-informed-DeepONet-framework}, the architecture consists of two parallel networks: a branch network and a trunk network. The branch network processes discretized input functions that represent the boundary conditions, initial conditions, parameters, or other forcing terms of the SWEs. These input functions are sampled at discrete points and encoded into a high-dimensional feature representation through the branch network. The trunk network, operating in parallel to the branch network, processes the spatiotemporal coordinates $(x, y, t)$ at which the solution is to be evaluated. For steady state problems in the present study, the input is only the spatial coordinates $(x, y)$. The outputs of the branch and trunk networks are then combined through an inner product (or more generally, a dot product) to produce the output of the operator. In the case of 2D hydraulics modeling, the output $\mathbf{q} = (h, u, v)$ consists of the water depth $h$ and the two velocity components ($u$ and $v$) at the spatial coordinates. Once trained, the model can predict the flow field at any given spatiotemporal coordinates without being constrained to a fixed computational mesh.

The training of the model (weights and biases in the branch and trunk networks) is carried out by minimizing the total loss function ($\mathcal{L}_{total}$) consisting of four loss terms: data loss ($\mathcal{L}_{data}$), physics loss ($\mathcal{L}_{PDEs}$), boundary conditions and initial conditions loss ($\mathcal{L}_{BC/IC}$), and other losses ($\mathcal{L}_{other}$); details are explained below. 

In the purely data-driven DeepONet mode, termed ``SWE-DeepONet'' in this work, only the data loss function $\mathcal{L}_{data}$ is minimized. Data loss measures the discrepancy between the predicted and the true values, which can be expressed as the mean square error (MSE):
\begin{equation}
    \mathcal{L}_{data} = \frac{1}{N_{data}} \sum_{i=1}^{N_{data}} \left( \hat{\mathbf{q}}_i - \mathbf{q}_i \right)^2
\end{equation}
where $\hat{\mathbf{q}}_i$ is the predicted value, $\mathbf{q}_i$ is the true value, and $N_{data}$ is the number of data points. The true value $\mathbf{q}_i$ can be generated by running physics-based numerical models such as SRH-2D or HEC-RAS 2D.  

In the physics-informed DeepONet mode, termed ``PI-SWE-DeepONet'' in this work, the flow physics governed by the SWEs is additionally incorporated into the training process by adding the physics loss function $\mathcal{L}_{PDEs}$. This loss function measures the violation of the governing equations and is computed as the summation of PDE residuals at each PDE point. At each point $i$, the PDE residual is a weighted average of the residuals for the continuity equation and the two momentum equations. The physics loss function is expressed as:
\begin{equation}\label{eqn:PDE_loss_components}
    \mathcal{L}_{PDEs} = \frac{1}{N_{PDEs}} \sum_{i=1}^{N_{PDEs}} \left[ w_c e_{cty,i}^2 + w_x e_{x\text{-momentum},i}^2 + w_y e_{y\text{-momentum},i}^2 \right]
\end{equation}
where $N_{PDEs}$ is the number of PDE points, and $e_{cty,i}$, $e_{x\text{-momentum},i}$, and $e_{y\text{-momentum},i}$ are the residuals of the three governing equations at point $i$. The residuals are computed by substituting the predicted flow field into each governing equation, with all partial derivatives evaluated using automatic differentiation (AD) of the neural networks \cite{Paszke2017Automatic,Baydin2018Automatic}. Weighting factors, $w_c$, $w_x$, and $w_y$ are for the continuity, $x$-momentum, and $y$-momentum equations, respectively. They designate the relative importance of the three equations in the physics loss function. 

The enforced 2D SWEs are in the non-conservative form and may be expressed as:
\begin{equation}\label{eqn:cty}
    \frac{\partial h}{\partial t} + \frac{\partial h u}{\partial x} + \frac{\partial h v}{\partial y} =0
\end{equation}
\begin{equation}\label{eqn:momentum_x}
    \frac{\partial u}{\partial t} + u\frac{\partial u }{\partial x} + v \frac{\partial u }{\partial y} = -g\frac{\partial \left(h+z_b\right)}{\partial x}-\frac{\tau_{bx}}{\rho} 
\end{equation}
\begin{equation}\label{eqn:momentum_y}
    \frac{\partial v}{\partial t} + u \frac{\partial v}{\partial x} + v \frac{\partial v}{\partial y} =  -g\frac{\partial \left(h+z_b\right)}{\partial y}-\frac{\tau_{by}}{\rho} 
\end{equation}
where $u$ and $v$ are the depth-averaged flow velocities in $x$ and $y$ directions, respectively; $h$ is the water depth; $z_b$ is the bed elevation;  $\tau_{bx}$ and $\tau_{by}$ are the bed shear stresses in $x$ and $y$ directions, respectively; $g$ is the gravitational acceleration; $\rho$ is the water density. The bed shear stresses are computed using the Manning's formula: $\tau_{bx} = \rho g n^2 u |\mathbf{u}| / h^{4/3}$ and $\tau_{by} = \rho g n^2 v |\mathbf{u}| / h^{4/3}$, where $n$ is the Manning's roughness coefficient and $\mathbf{u}$ is the velocity vector. Details about the governing equations and the physical meanings of all terms are referred to \citeA{lai2010two}.

In this study, the non-conservative form of the 2D SWEs is adopted, versus the conservative form, because it produces simpler automatic differentiation graphs and reduces residual imbalance during training within the physics-informed DeepONet framework. Deep-learning surrogate models generally do not enforce conservation laws on a control volume by computing fluxes across control-volume boundaries. Instead, they enforce conservation through pointwise PDE residuals. This choice of using non-conservative form is also consistent with prior physics-informed SWE studies \cite{bihlo2022} and is physically valid because the supervised SRH-2D data inherently enforce conservation at the solution level.

Additional optional losses may be added besides the data loss and the physics loss, although they are not used in this work. For example, boundary conditions (BC) and initial conditions (IC) can be enforced through the combined loss of $\mathcal{L}_{BC/IC}$. The computation of BC and IC loss needs the information of the boundary conditions and initial conditions. Other losses such as the regularization loss can also be added to the total loss function. In deep learning, regularization loss is typically a penalty term on the magnitude of the neural network weights and biases (e.g., L1 or L2 regularization), which helps prevent overfitting by discouraging large parameter values. These losses are optional and they are omitted for the examples in this work. 

The total loss function is then expressed as a weighted sum of all loss components:
\begin{equation}
    \mathcal{L}_{total} = \lambda_{data} \mathcal{L}_{data} + \lambda_{PDEs} \mathcal{L}_{PDEs} + \lambda_{BC/IC} \mathcal{L}_{BC/IC} + \lambda_{other} \mathcal{L}_{other}
\end{equation}
where the weights $\lambda_{data}$, $\lambda_{PDEs}$, $\lambda_{BC/IC}$, and $\lambda_{other}$ are hyperparameters that measure the relative importance of corresponding loss components. 

\subsection{Input Configurations}\label{subsec:input-configuration}
The input to the branch network is flexible and can accommodate various types of boundary conditions and forcing functions depending on the specific application. In the example shown in Figure~\ref{fig:physics-informed-DeepONet-framework}, the input is a hydrograph function represented as a time series of discharge values at discrete time points for an unsteady flow simulation. However, the framework can readily handle other input configurations. For steady-state problems, the input can be simplified to discharge values at different model inlets, eliminating the temporal dimension. This flexibility makes the framework adaptable to diverse hydraulic modeling scenarios, from constant discharges (steady-state) to complex unsteady open channel flows with multiple inflow and exit boundaries. 

In this study, the algorithms and workflow presented are implemented in the open-source Python package HydroNet, which uses PyTorch for deep learning model development and training. The neural network architectures, including both the branch and trunk networks of DeepONet, are built using PyTorch's neural network modules, enabling automatic differentiation for computing PDE residuals in physics-informed training \cite{Paszke2017Automatic,Paszke2019PyTorch}. The framework also leverages pyHMT2D, a Python interface for the SRH-2D hydraulic model, to process simulation results, extract flow field data from result files, and generate collocation points for physics-informed training. The implementation supports both CPU and GPU computing. GPU acceleration significantly reduces training time for large-scale problems, while CPU execution provides flexibility for systems without GPU access. 

\section{Application Example: Sacramento River}\label{sec:application-example-sacramento-river}

The methodology and tools developed in this study are applied to a real-world case: a section of the Sacramento River, California. The Sacramento River is one of the major rivers in California, USA, and has historically been subject to complex flow dynamics due to its confluence with the San Joaquin River and the intricate channel geometry in the delta region. In this study, a section of the river near the Fremont Weir is simulated under a constant-discharge flow condition. Extensive hydraulic modeling studies have been conducted at this site in which both SRH-2D and 3D simulations were reported and verified with measurement data \cite{lai2D3DFlow2016}. 

\subsection{Domain and Mesh}\label{subsec:sacramento-river-domain-and-mesh}

\begin{figure}[htp]
    \centering
    \includegraphics[width=1.0\textwidth]{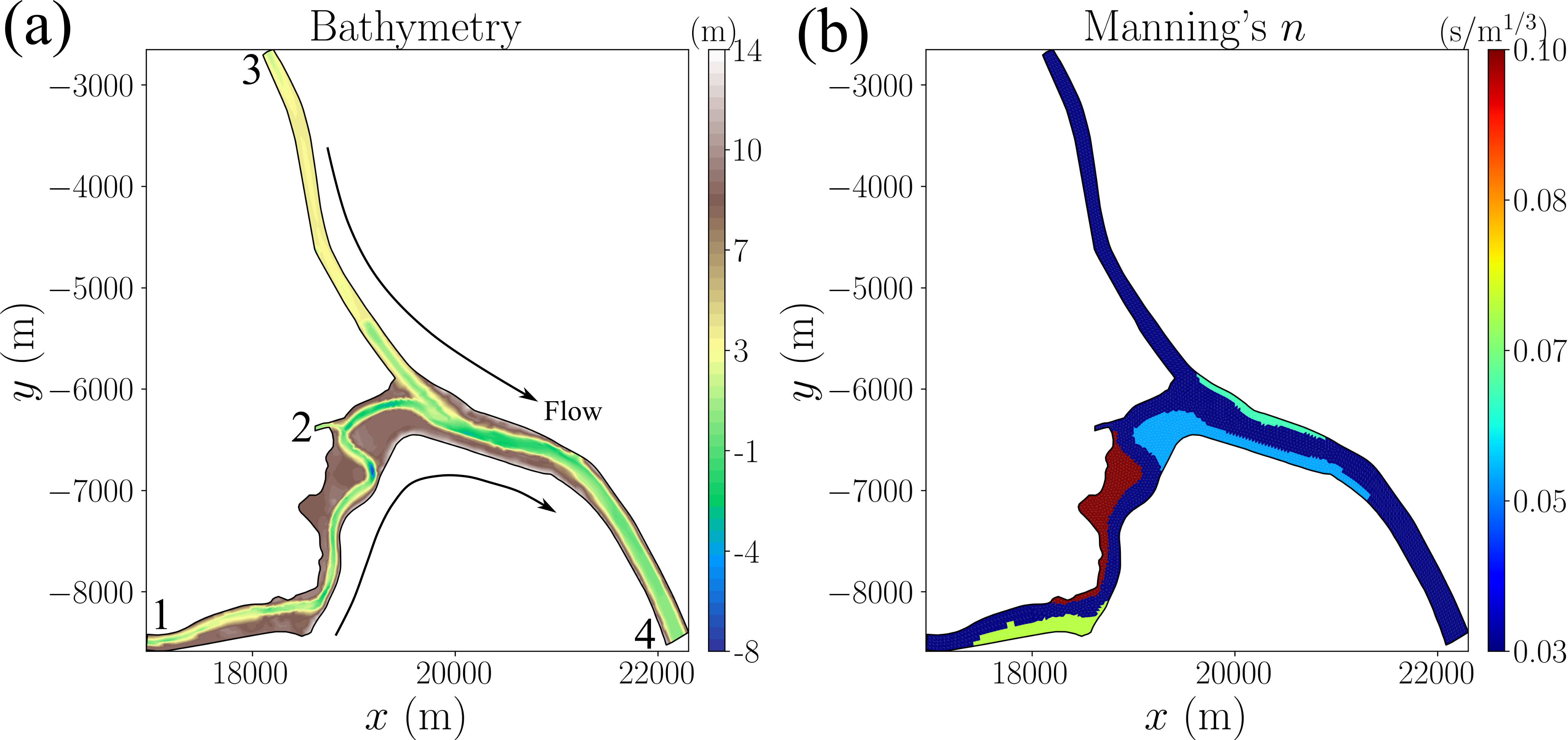}
    \caption{The Sacramento River model domain: (a) bathymetry with the inlet and outlet locations marked, (b) Manning's $n$ values within the domain.}
    \label{fig:sacramento-river-domain}
\end{figure}

The model domain in this study is shown in Figure~\ref{fig:sacramento-river-domain}(a), which includes the confluence of the Feather River and Sacramento River with an additional inlet flow from the Sacramento Slough at Karnak. The modeled domain is about 6 km long and 5 km wide. The river width ranges from 150 to 250 meters. The bed elevation ranges from about -8 meters at the thalweg of the left channel to 14 meters along the banks near confluence. The modeled domain has four boundaries labeled as 1, 2, 3, and 4 in the figure. Boundaries 1 to 3 are inlets with specified flow discharges. Boundary 4 is an outlet modeled with a stage-discharge rating curve. The domain is divided into five roughness zones with Manning's $n$ values ranging from 0.03 in the main channel to 0.1 in the floodplain (Figure~\ref{fig:sacramento-river-domain}(b)).

\begin{figure}[htp]
    \centering
    \includegraphics[width=1.0\textwidth]{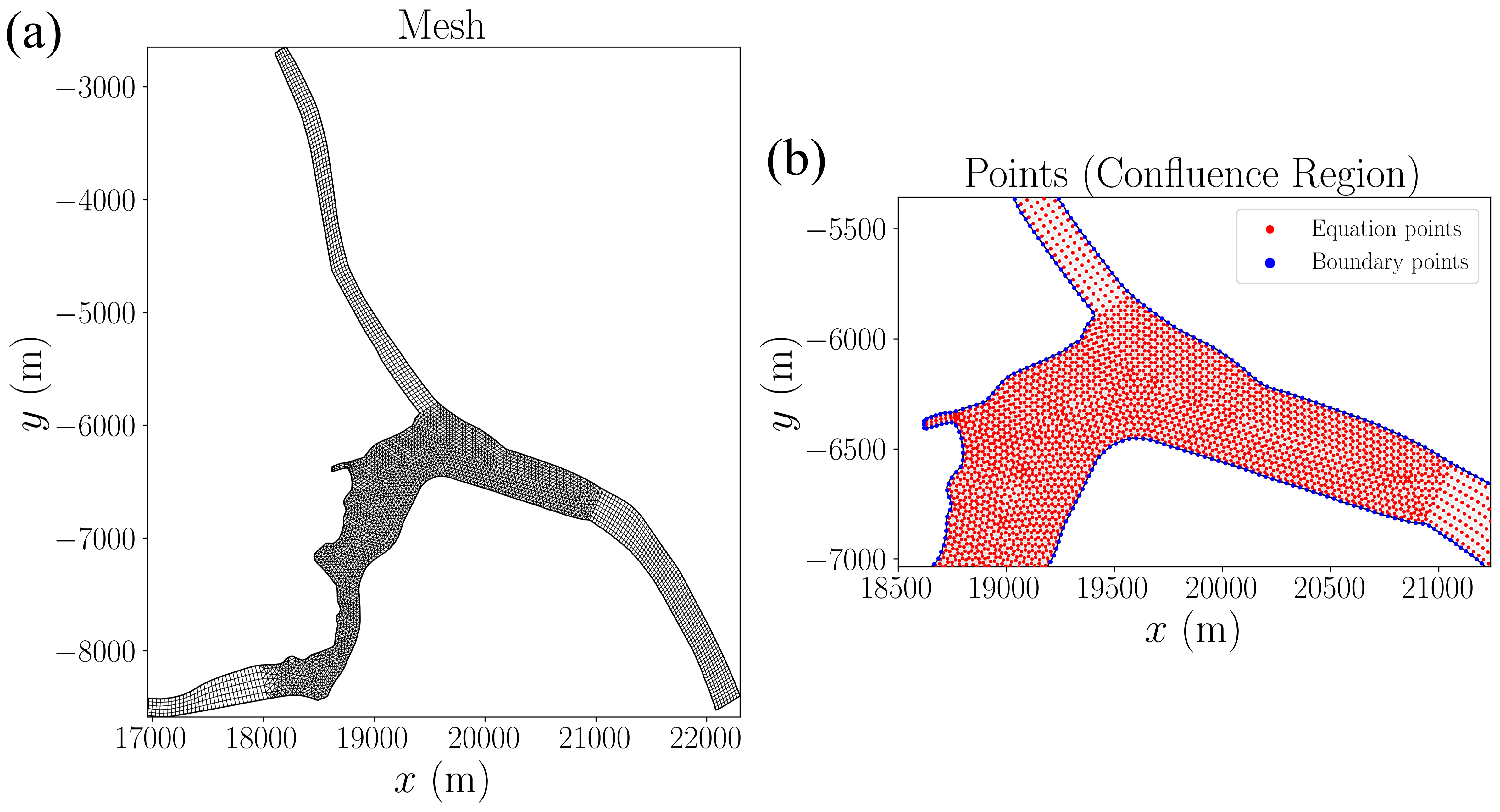}
    \caption{Mesh and collocation points for the Sacramento River case: (a) SRH-2D mesh, (b) Collocation points in the confluence area for the PI-SWE-DeepONet model.}
    \label{fig:sacramento-river-mesh-points}
\end{figure}

A 2D mesh is generated for use by SRH-2D and shown in Figure~\ref{fig:sacramento-river-mesh-points}(a); it has 5,036 mixed triangle-quad cells. For the DeepONet surrogate models, collocation points are additionally needed on which the PDE residuals and loss functions due to boundary conditions are computed. In this study, a Python tool was developed to generate the collocation points based on the SRH-2D mesh. For PDE residual points, the centers of all 2D mesh cells are used though a different set of points may also be used. For the boundary condition loss function, all boundary edges of the mesh are traced to generate the points. The PDE and boundary collocation points in the zoom-in area of the confluence are displayed in Figure~\ref{fig:sacramento-river-mesh-points}(b). Note that the collocation points are the same as the mesh centers in this study as it was found that the approach is adequate and sufficient, at least for the case simulated.

\subsection{Training, Validation, and Test Data Sets} \label{subsec:data-generation}

\begin{figure}[htp]
    \centering
    \includegraphics[width=0.9\textwidth]{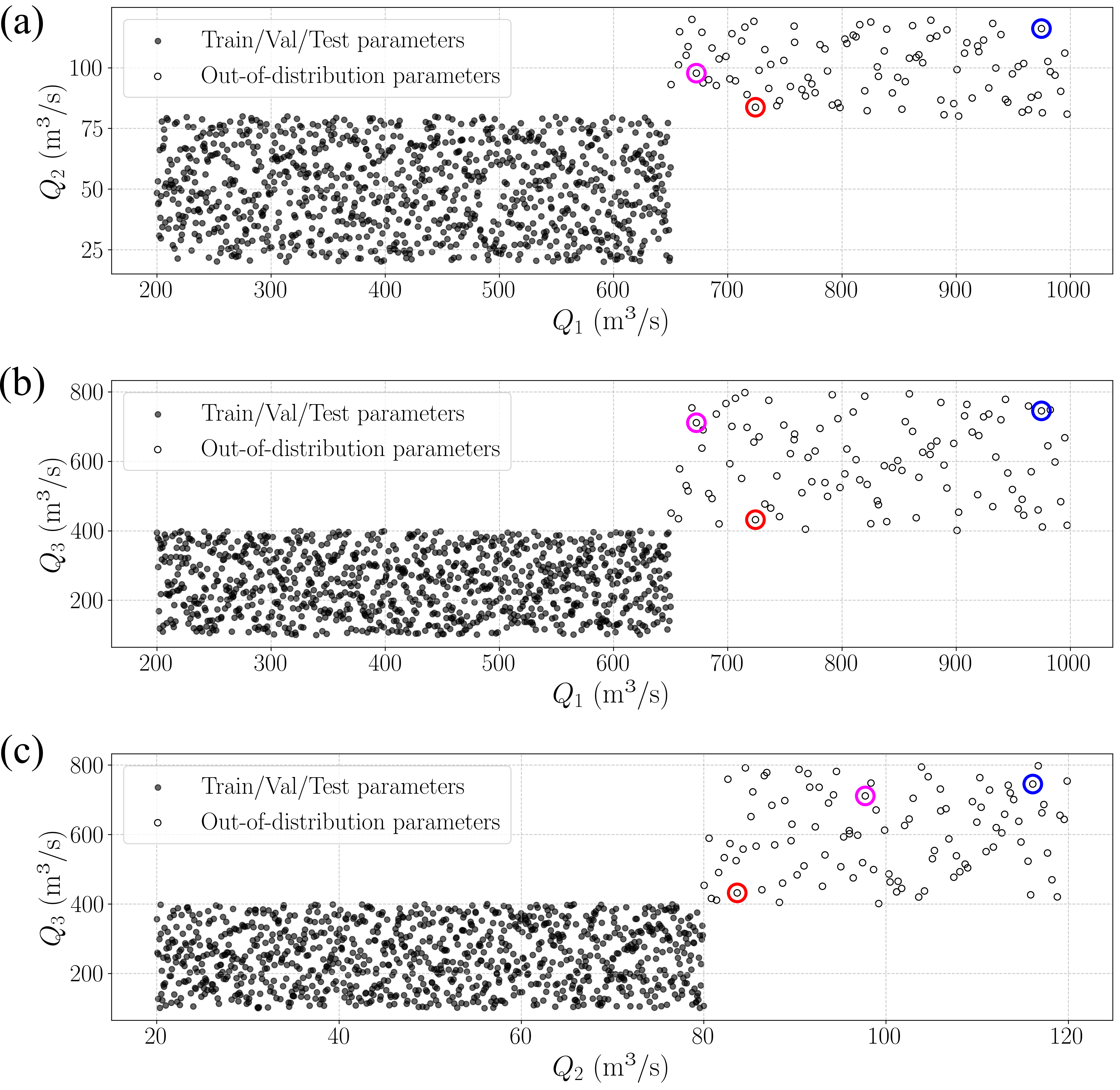}
    \caption{Sampled inlet discharge values at selected inlet pairs: (a) $Q_1$ vs $Q_2$, (b) $Q_1$ vs $Q_3$, (c) $Q_2$ vs $Q_3$. The training/validation/test and out-of-distribution parameters are plotted in solid and open circles, respectively. Three out-of-distribution parameter sets with minimum, average, and maximum Wasserstein distances are circled in red, magenta, and blue, respectively.}
    \label{fig:sacramento-river-inlet-discharge-pairs}
\end{figure}

The training, validation, and test data sets were generated through a systematic Monte Carlo parameter sampling and SRH-2D simulation workflow. The varying parameters, through the branch inputs to the DeepONet model, are the discharges at the three inlets, noted as $Q_1$, $Q_2$, and $Q_3$. Parameter samples were generated using Latin Hypercube Sampling (LHS) for three inlet discharge boundary conditions with uniform distributions over the ranges [200, 650] m$^3$/s, [20, 80] m$^3$/s, and [100, 400] m$^3$/s, respectively. A total of 1,000 parameter combinations were sampled. The solid black circles in Figure~\ref{fig:sacramento-river-inlet-discharge-pairs} show the sampled inlet discharge values at selected inlet pairs.

For each parameter combination, an SRH-2D simulation was first performed, producing flow field solutions (water depth $h$ and velocity components $u$ and $v$) at all mesh cell centers. The simulation results were then extracted and postprocessed to prepare the input data needed for DeepONet training. The 1,000 simulation cases were randomly split into training (70\%), validation (20\%), and test (10\%) data sets using a stratified random split. For each split, the data were organized into branch inputs (the three inlet discharge values), trunk inputs (spatial coordinates $x$ and $y$ at cell centers), outputs (water depth and velocity components), and PDE data (PDE collocation point coordinates, bed elevation, Manning's $n$, and bed slope components) required for physics-informed training.

Feature variables in machine learning require appropriate normalization to ensure stable optimization and effective convergence. Different normalization strategies are adopted to reflect the distinct roles of the variables in the learning process. The branch inputs, namely the inlet discharge values $Q_1$, $Q_2$, and $Q_3$, are normalized using z-score normalization so that each discharge is centered and scaled by its standard deviation across all training cases, preventing dominance by larger-magnitude inputs. The trunk inputs, consisting of the spatial coordinates $x$ and $y$, are normalized using min–max scaling to map each coordinate to a bounded interval between zero and one, which preserves the geometric structure of the domain and improves numerical conditioning for evaluating spatial derivatives in the physics-informed loss. The outputs, including water depth $h$ and velocity components $u$ and $v$, are normalized using z-score normalization based on statistics computed from all SRH-2D simulations, ensuring comparable gradient magnitudes across variables with different physical units and dynamic ranges. Together, these choices promote stable training while maintaining the physical consistency of both the input coordinates and the hydraulic state variables.

For the PDE loss, the normalization was done differently. The predicted flow variables $h$, $u$, and $v$ from the DeepONet models were first denormalized. Then, the residuals of the continuity equation and the $x$ and $y$ momentum equations, were computed using the denormalized values. The residuals were then normalized using the maximum length scale $L$ and velocity scale $U$ of the Sacramento River cases. The length scale $L$ was the maximum water depth. Specifically, the continuity equation residual was normalized by $U$, and the $x$ and $y$ momentum equation residuals were normalized by $U^2/L$. 

\subsection{Out-of-Distribution Cases}\label{subsec:out-of-distribution-cases}

Out-of-distribution cases were generated to evaluate the generalizability of the SWE-DeepONet and PI-SWE-DeepONet models. These are the cases not in the training data distribution. The out-of-distribution parameter sets were selected by extending the inlet discharge ranges beyond those used in the training data. Specifically, the training data used inlet discharge ranges of [200, 650] m$^3$/s, [20, 80] m$^3$/s, and [100, 400] m$^3$/s for $Q_1$, $Q_2$, and $Q_3$, respectively. In contrast, the out-of-distribution cases were sampled from extended uniform distributions over the ranges [650, 1000] m$^3$/s, [80, 120] m$^3$/s, and [400, 800] m$^3$/s. A total of 100 out-of-distribution parameter combinations were generated and shown as the open black circles in Fig.~\ref{fig:sacramento-river-inlet-discharge-pairs}. It can be observed that the out-of-distribution inlet discharge values extend beyond the training data distribution (solid black circles). For each out-of-distribution parameter combination, a constant-discharge SRH-2D model run was performed to produce reference solutions for model evaluation. 

\begin{figure}[htp]
    \centering
    \includegraphics[width=0.9\textwidth]{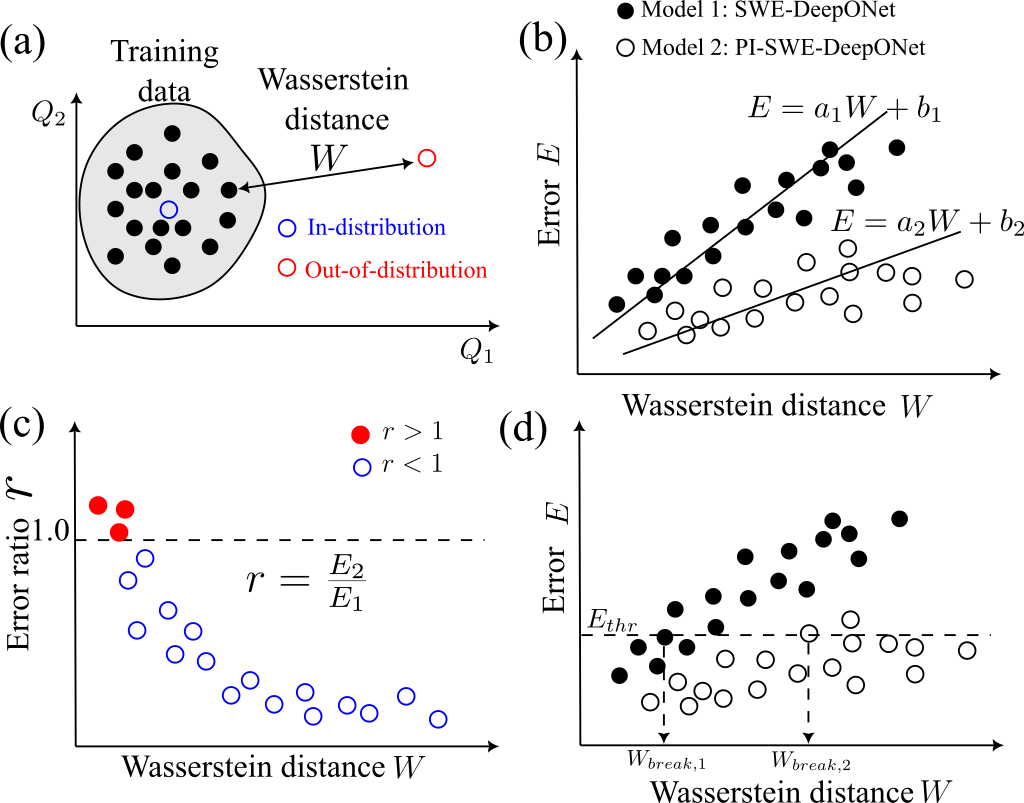}
    \caption{Evaluation metrics for the out-of-distribution cases: (a) Wasserstein distance, (b) error sensitivity, (c) error ratio, (d) model breakdown distance.}
    \label{fig:evaluation-metrics}
\end{figure}

For evaluation, the Wasserstein distance $W$ was computed for each out-of-distribution case \cite{kantorovich1958translocation,villani2009optimal}; it measures how far each out-of-distribution parameter set is from the training data. The Wasserstein distance is defined as the L1 (Manhattan) distance to the nearest training sample in the normalized parameter space, i.e., $W(\mathbf{Q}) = |\mathbf{Q} - \mathbf{Q}_{\mathrm{closest}}|_1$, where $\mathbf{Q}$ represents the normalized inlet discharge parameter vector [$Q_1$, $Q_2$, $Q_3$] and $\mathbf{Q}_{\mathrm{closest}}$ is the closest training sample (see Fig.~\ref{fig:evaluation-metrics}a). Three representative out-of-distribution cases were identified based on their Wasserstein distances: cases with minimum distance (closest to training data), average distance, and maximum distance (farthest from training data). In Fig.~\ref{fig:sacramento-river-inlet-discharge-pairs}, these three cases are highlighted with colored circles: red for minimum distance, magenta for average distance, and blue for maximum distance. These three cases will be referenced and analyzed in detail in later sections to evaluate the model performance across different levels of extrapolation. 

To quantify the performance of the DeepONet models on out-of-distribution cases, three evaluation metrics are computed and shown in Fig.~\ref{fig:evaluation-metrics}:
\begin{itemize}
    \item The error sensitivity $a$: It is computed as the slope of the linear regression ($E=aW+b$) between Wasserstein distance and model prediction error. It characterizes how rapidly prediction errors increase with distance from the training data (Fig.~\ref{fig:evaluation-metrics}b). A steeper slope indicates higher sensitivity, indicating that the model's performance degrades more quickly as it extrapolates beyond the training distribution.
    \item The error ratio $r=E_2/E_1$: Here $E_1$ and $E_2$ are the model prediction errors for the SWE-DeepONet and PI-SWE-DeepONet models, respectively (Fig.~\ref{fig:evaluation-metrics}c). A ratio of $r<1$ indicates that the physics constraint is effective in reducing the prediction error, while a ratio of $r>1$ indicates the opposite. The fraction of $r<1$, noted as $f_{r<1}$, is the fraction of cases where the physics constraint improves the prediction.
    \item The model breakdown distance: It is defined as the minimum Wasserstein distance at which the prediction error exceeds a specified threshold $E_{thr}$ (e.g., normalized RMSE > 0.5), providing a quantitative measure of the model's extrapolation limit (Fig.~\ref{fig:evaluation-metrics}d). 
\end{itemize}
These metrics collectively provide a comprehensive assessment of model generalizability and robustness when applied to scenarios outside the training data distribution. 

\subsection{Training, hyperparameters, and weight adjustment}

Both SWE-DeepONet and PI-SWE-DeepONet models were trained for 1,000 epochs using the Adam optimizer with an initial learning rate of 0.001 and weight decay of $10^{-5}$ for L2 regularization. The learning rate was dynamically adjusted using a scheduler that reduces the learning rate by a factor of 0.5 when the validation loss plateaus for 10 consecutive epochs, with a minimum learning rate of $10^{-6}$. The network architecture consists of branch and trunk networks, each with six fully connected hidden layers of sizes [64, 128, 256, 256, 128, 129] neurons. The branch network processes the three inlet discharge inputs, while the trunk network processes the spatial coordinates $(x, y)$. Training was performed with mini-batch over simulation cases and a batch size of four was used. 

Activation functions in the branch and trunk networks play a crucial role in the performance of the DeepONet models. Both ReLU and $\tanh$ activation functions were explored. For the purely data-driven SWE-DeepONet, ReLU activation function was employed due to their computational efficiency and favorable gradient propagation properties, which help mitigate potential vanishing gradients when training relies solely on data mismatch. For the PI-SWE-DeepONet, $\tanh$ activation function was adopted because its smooth and continuously differentiable nature offers a key advantage for stable and accurate evaluation of PDE residuals via automatic differentiation. It is noted that $\tanh$ function can exhibit gradient saturation for large input values. However, this effect is alleviated through nondimensionalization of inputs and outputs and the use of moderate network depth. In contrast, the ReLU activation function is not differentiable everywhere and has zero higher-order derivatives, which can hinder physics-informed model training.

For the PI-SWE-DeepONet model, the weights need to be given between data loss and physics loss. In this study, adaptive loss balancing was employed so that the relative weights between the two are determined automatically at an interval of 10 epochs throughout training. The approach performed well in balancing the two different loss scales and stable convergence was ensured. The target weights were pre-specified for both data and physics losses and the adaptive balancing mechanism adjusted them based on the relative magnitudes of the loss components during training. To smooth the loss magnitude estimates used for the weight adjustment, exponential moving averaging with a smoothing factor of 0.1 was applied to both data loss and PDE loss magnitudes, which effectively reduced noise and provided stable weight updates throughout the training process.

The target loss weights for the data and PDE losses, i.e., $\lambda_{data}$ and $\lambda_{PDEs}$, can have an important impact on the model training and prediction performance. In this study, three different sets of target loss weights [$\lambda_{data}$, $\lambda_{PDEs}$] were tried: [2.0, 1.0], [1.0, 1.0], and [1.0, 2.0]. Indeed, the ratio $\lambda_{data}/\lambda_{PDEs}$ measures the relative importance of data fitting and physics consistency in the trained surrogate model. No rule of thumb has been developed yet for the optimal ratio. In general, it depends on how much emphasis a user prefers to place on the physics consistency compared to the data fitting. In this study, the ratio was selected such that the PI-SWE-DeepONet model performance on in-distribution test cases is close to the performance of the SWE-DeepONet model. With this criterion, the target weights for data and PDE losses were found to be [2.0, 1.0]. A comparative study on the impact of the loss weights is presented in the Supplemental Information.

\section{Results and Analyses}\label{sect:results}

\subsection{Training and Validation Loss Histories}\label{subsec:training-validation-losses}
\begin{figure}[htp]
    \centering
    \includegraphics[width=0.9\textwidth]{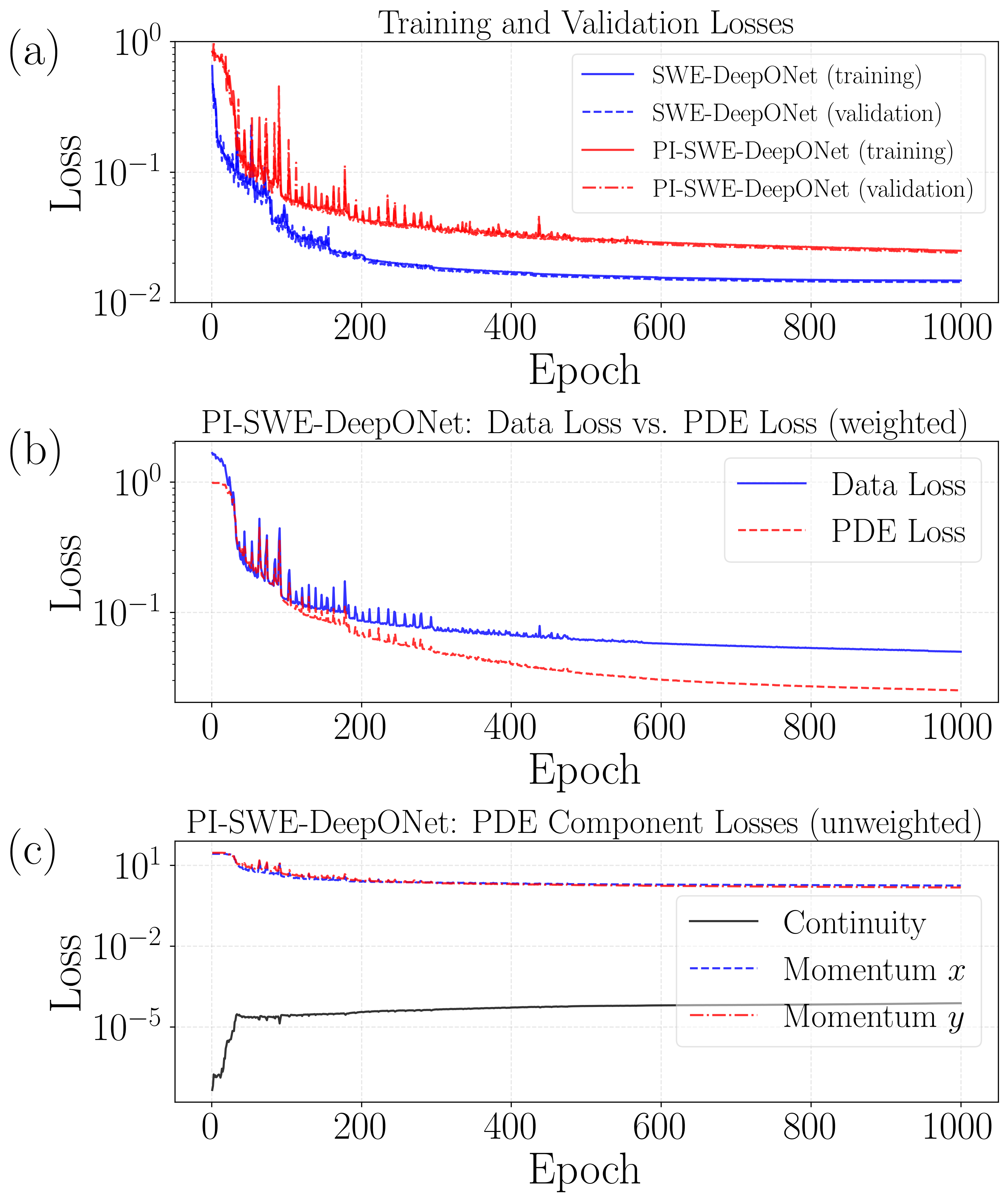}
    \caption{The training histories of the SWE-DeepONet and PI-SWE-DeepONet models for the Sacramento River case: (a) the training and validation losses for both models, (b) the weighted data and PDE losses for the PI-SWE-DeepONet model, (c) the unweighted PDE component losses for the PI-SWE-DeepONet model.}
    \label{fig:training-history}
\end{figure}

The loss histories for both SWE-DeepONet and PI-SWE-DeepONet models convey important messages and are shown in Fig.~\ref{fig:training-history}. Both models achieved convergence during training with no signs of overfitting, as evidenced by the parallel trends between training and validation losses throughout the 1,000 epochs. Figure~\ref{fig:training-history}(a) compares the training and validation losses for both models. For a fair comparison, the validation loss of PI-SWE-DeepONet shown in this subfigure contains only the data loss component, excluding the PDE physics loss. Notably, the final validation loss of PI-SWE-DeepONet is higher than that of SWE-DeepONet. This difference reflects the important trade-off inherent in physics-informed training: by incorporating the physics constraint, the model must simultaneously satisfy both the training data and the governing PDEs, which can lead to an increase in data fitting error. This cost in data loss, however, is offset by the improved physical consistency and enhanced generalization capability demonstrated in the out-of-distribution cases to be discussed later. 

Figure~\ref{fig:training-history}(b) shows the weighted data loss and PDE loss components for the PI-SWE-DeepONet model throughout training. The weighted losses gradually converge to the target loss ratio of 2:1, demonstrating the effectiveness of the adaptive loss balancing mechanism. The exponential moving averaging applied to loss magnitude estimates reduces noise in the weight adjustment process, leading to training stability and smooth convergence. 

Figure~\ref{fig:training-history}(c) displays the unweighted PDE loss components, showing the individual residuals from the continuity equation and the $x$ and $y$ momentum equations. For the Sacramento River case studied, the continuity equation residual shows orders of magnitude smaller than the momentum residuals, which decrease steadily during training. This indicates that the model is learning to satisfy the momentum equations across the domain and the continuity equation plays a minor role for the training. This significant disparity in mass balance and momentum residuals is the reason that they were not weighted in the computation of the PDE loss with the present case, i.e., $w_c$, $w_x$ and $w_y$ in Eq.~\ref{eqn:PDE_loss_components} all had a value of 1.0. If they were weighted to have comparable contribution to the PDE loss, the training process will try to reduce the already very small continuity residual even more, which does not make sense.

\subsection{Model Performance with Test Cases}

The performance of both SWE-DeepONet and PI-SWE-DeepONet models with the test cases is evaluated by comparing their predictions against SRH-2D simulation results. Test cases are ``in-distribution'', meaning their inlet discharges are within the same distribution as the training data. The test case data were never seen by the ML models during training. Thus, the performance with test cases provides an unbiased assessment of model accuracy within the training parameter distribution.

\begin{figure}[htp]
        \centering
        \includegraphics[width=1\textwidth]{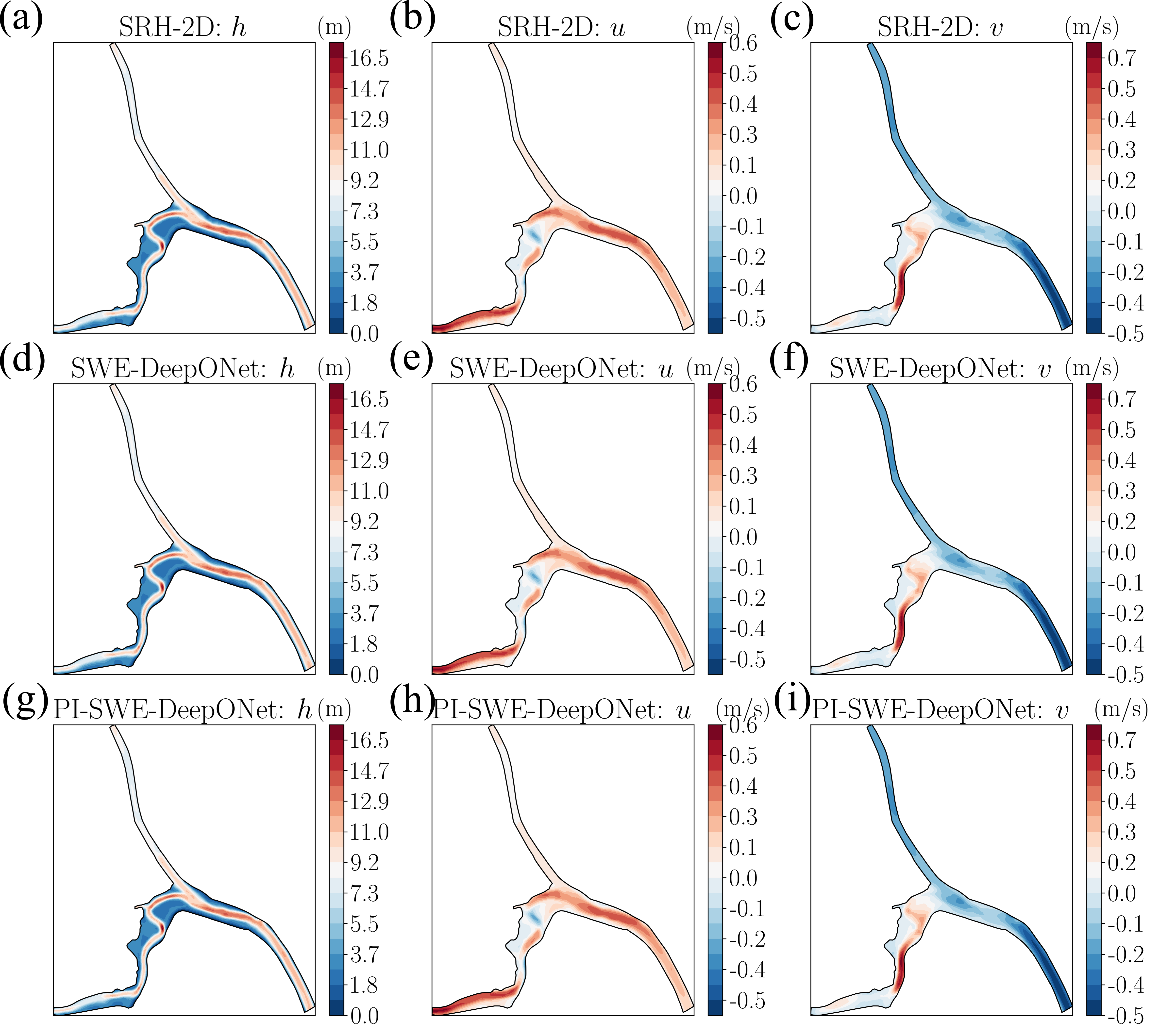}
        \caption{Comparison of the SWE-DeepONet and PI-SWE-DeepONet models against SRH-2D result for one representative test case with the inlet discharges $Q_1$ = 582.8 m$^3$/s, $Q_2$=49.1 m$^3$/s, $Q_3$=258.4 m$^3$/s: (a) $h$ from SRH-2D, (b) $u$ from SRH-2D, (c) $v$ from SRH-2D, (d) $h$ from SWE-DeepONet, (e) $u$ from SWE-DeepONet, (f) $v$ from SWE-DeepONet, (g) $h$ from PI-SWE-DeepONet, (h) $u$ from PI-SWE-DeepONet, (i) $v$ from PI-SWE-DeepONet.}
        \label{fig:compare-test-case-results}
    \end{figure}

\begin{figure}[htp]
    \centering
    \includegraphics[width=0.5\textwidth]{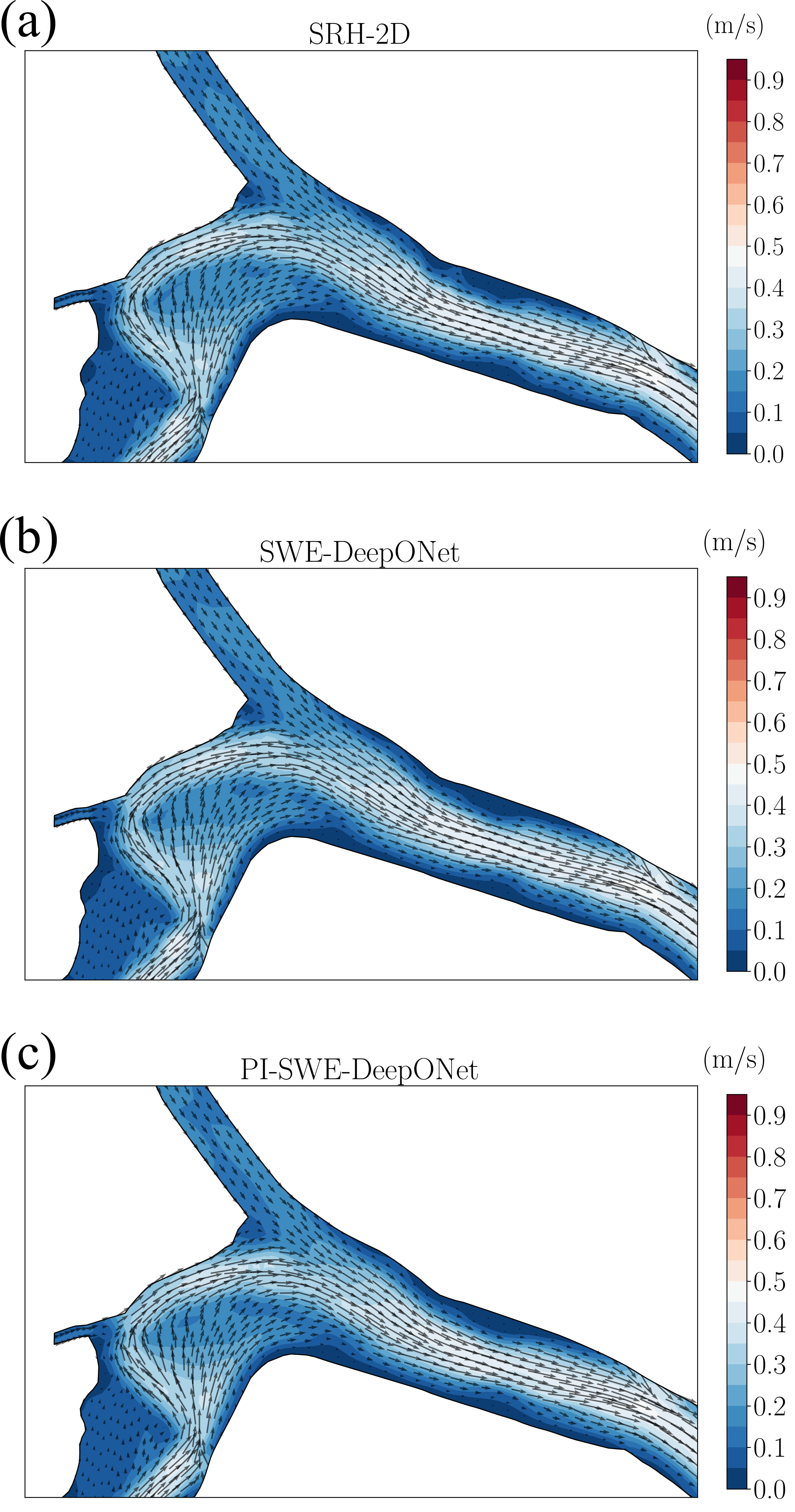}
    \caption{Comparison of velocity vector fields for one representative test case: (a) SRH-2D, (b) SWE-DeepONet, (c) PI-SWE-DeepONet.}
    \label{fig:compare-test-case-results-vector}
\end{figure}

Figure~\ref{fig:compare-test-case-results} compares the spatial distributions of water depth $h$, velocity components $u$ and $v$ from the two surrogate models and SRH-2D for one representative case. The inlet discharges for this case are $Q_1$ = 582.8 m$^3$/s, $Q_2$=49.1 m$^3$/s, $Q_3$=258.4 m$^3$/s, respectively. Figure~\ref{fig:compare-test-case-results-vector} is the zoomed-in views of the velocity vector fields. Both surrogate models capture the overall flow patterns well, including the complex flow dynamics at the confluence where the two main rivers meet. The water depth distributions show notable differences in the magnitude ranges: SWE-DeepONet reproduces the SRH-2D water depth range of [0, 18] m, while PI-SWE-DeepONet predicts a larger range of [0, 19.6] m, which indicates an overprediction of the maximum water depth. This difference is consistent with the early loss history analysis and attributes to the trade-off needed to accommodate the physics-informed training. Use of additional physics constraint may lead to a reduction in data fitting accuracy. Despite the difference in magnitude ranges, similar spatial patterns in water depth distribution are captured by both surrogate models, indicating the difference is limited to small isolated regions. The velocity fields are also predicted well by both surrogate models, particularly in the main channels. Both models accurately reproduce the flow direction and magnitude patterns observed in the SRH-2D results. The velocity vectors near the confluence demonstrate that both models have captured the flow features, though some differences exist in details.

\begin{figure}[htp]
    \centering
    \includegraphics[width=1\textwidth]{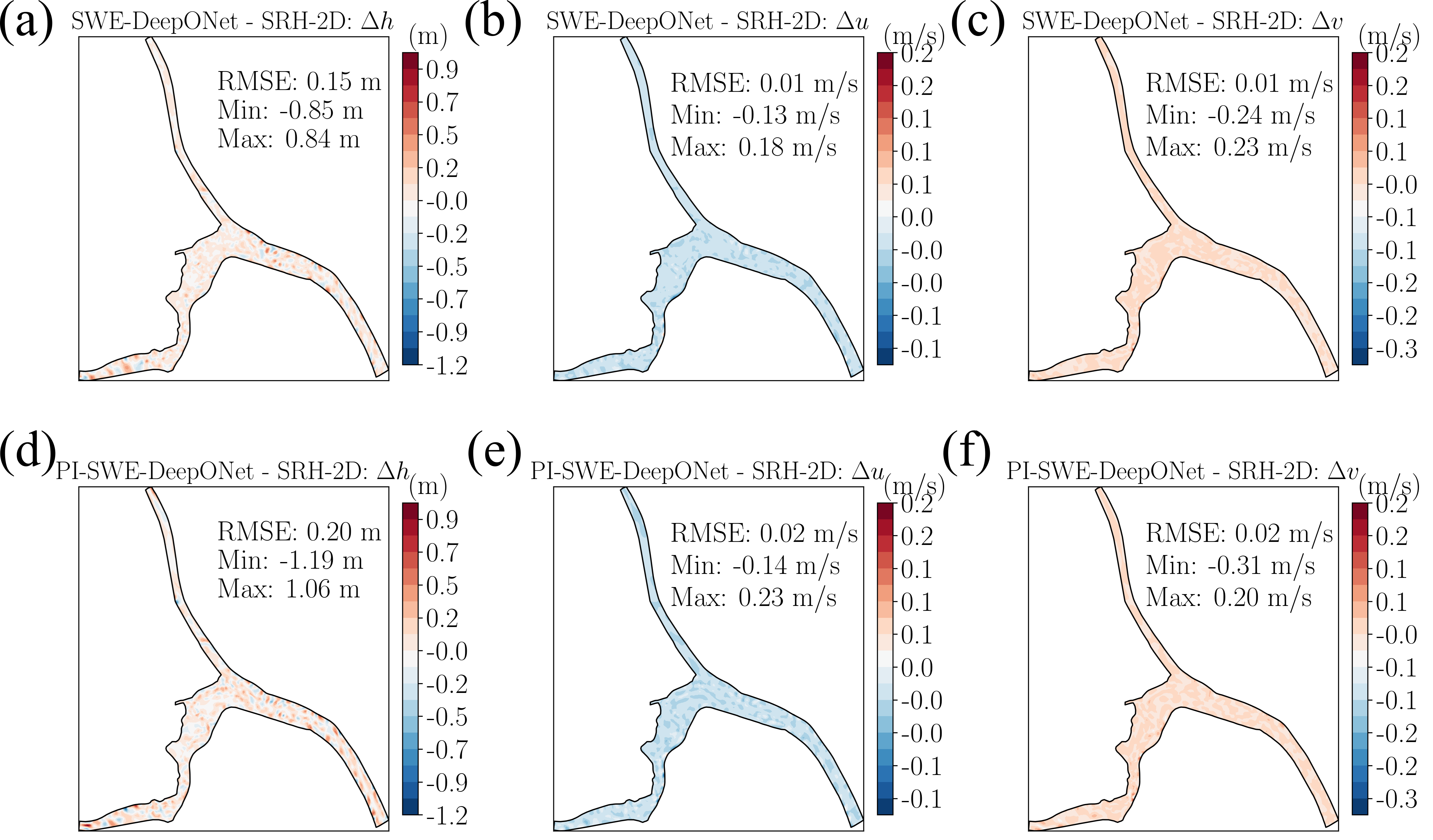}
    \caption{The differences between surrogate models and SRH-2D for one representative test case: (a) $\Delta h$ between SWE-DeepONet and SRH-2D, (b) $\Delta u$ between SWE-DeepONet and SRH-2D, (c) $\Delta v$ between SWE-DeepONet and SRH-2D, (d) $\Delta h$ between PI-SWE-DeepONet and SRH-2D, (e) $\Delta u$ between PI-SWE-DeepONet and SRH-2D, (f) $\Delta v$ between PI-SWE-DeepONet and SRH-2D.}
    \label{fig:compare-test-case-diff}
\end{figure}

For the same test case, Figure~\ref{fig:compare-test-case-diff} shows the difference (error) between the surrogate model predictions and the SRH-2D solution, i.e., error contour maps and statistical error metrics. It can be observed that the differences are generally small and well-distributed within the model domain and with no systematic bias patterns exist. The error statistics, including root mean square error (RMSE), minimum error, and maximum error, are displayed on each subfigure. For the water depth $h$ with an average value of about 5 m, both surrogate models achieved low RMSE values on the order of decimeters, demonstrating good accuracy in predicting water surface elevations. The velocity component errors ($\Delta u$ and $\Delta v$) are also relatively small, with RMSE values less than 0.02 m/s, indicating good agreement with SRH-2D solution. In general, PI-SWE-DeepONet has slightly larger, but consistent, errors in comparison with SWE-DeepONet. Again, this is consistent with the physics-constraint trade-off analysis discussed earlier. 

The quantitative performance metrics for all 100 test cases are computed and summarized in Table~\ref{tab:test-case-results}. Both SWE-DeepONet and PI-SWE-DeepONet models achieved good accuracy on all test cases within the training distribution. SWE-DeepONet in general has lower errors with all metrics than PI-SWE-DeepONet: an average normalized RMSE of 0.122 versus 0.157 for PI-SWE-DeepONet. For the individual flow variables, SWE-DeepONet achieved normalized RMSE values of 0.057, 0.156, and 0.130 for water depth $h$ and velocity components $u$ and $v$, respectively, while PI-SWE-DeepONet has shown corresponding values of 0.077, 0.200, and 0.167. In terms of physical units, SWE-DeepONet's denormalized RMSE values are 0.229 m for water depth and 0.024 m/s and 0.026 m/s for the $u$ and $v$ velocity components, respectively. PI-SWE-DeepONet shows denormalized RMSE values of 0.308 m for water depth and 0.031 m/s and 0.033 m/s for the velocity components. The higher errors in PI-SWE-DeepONet are again attributed to the inherent trade-off with added physics constraint. Despite this trade-off, both surrogate models demonstrate good performance in comparison with SRH-2D and model errors remain acceptably small. 

\begin{table}[htp]\label{tab:test_cases_performance}
    \centering
    \caption{Performance comparison between SWE-DeepONet and PI-SWE-DeepONet models with 100 test cases.}
    \label{tab:test-case-results}
    \renewcommand{\arraystretch}{1.5}
    \begin{tabular}{lccccccc}
        \toprule
        \multirow{2}{*}{Model} & \multirow{2}{2.5cm}{Average Normalized RMSE} & \multicolumn{3}{p{2.5cm}}{Normalized RMSE} & \multicolumn{3}{p{2.5cm}}{Denormalized RMSE} \\
        \cmidrule(lr){3-5} \cmidrule(lr){6-8}
        & & $h$ & $u$ & $v$ & $h$ (m) & $u$ (m/s) & $v$ (m/s) \\
        \midrule
        SWE-DeepONet & 0.122 & 0.057 & 0.156 & 0.130 & 0.229 & 0.024 & 0.026 \\
        PI-SWE-DeepONet & 0.157 & 0.077 & 0.200 & 0.167 & 0.308 & 0.031 & 0.033 \\
        \bottomrule
    \end{tabular}
    \renewcommand{\arraystretch}{1.0}
\end{table}

\subsection{Model Performance with Out-of-Distribution Cases}\label{subsec:out-of-distribution-results}

The performance on out-of-distribution cases is now presented to show the generalizability of trained surrogate models when scenarios are beyond the parameter range of the training data. Similar to the test cases in the previous section, 100 out-of-distribution cases were used and they were never seen by the surrogate models during the training. The discharge parameters of the out-of-distribution cases do not follow the same distribution as that of training data. 

\begin{figure}[htp]
        \centering
        \includegraphics[width=1\textwidth]{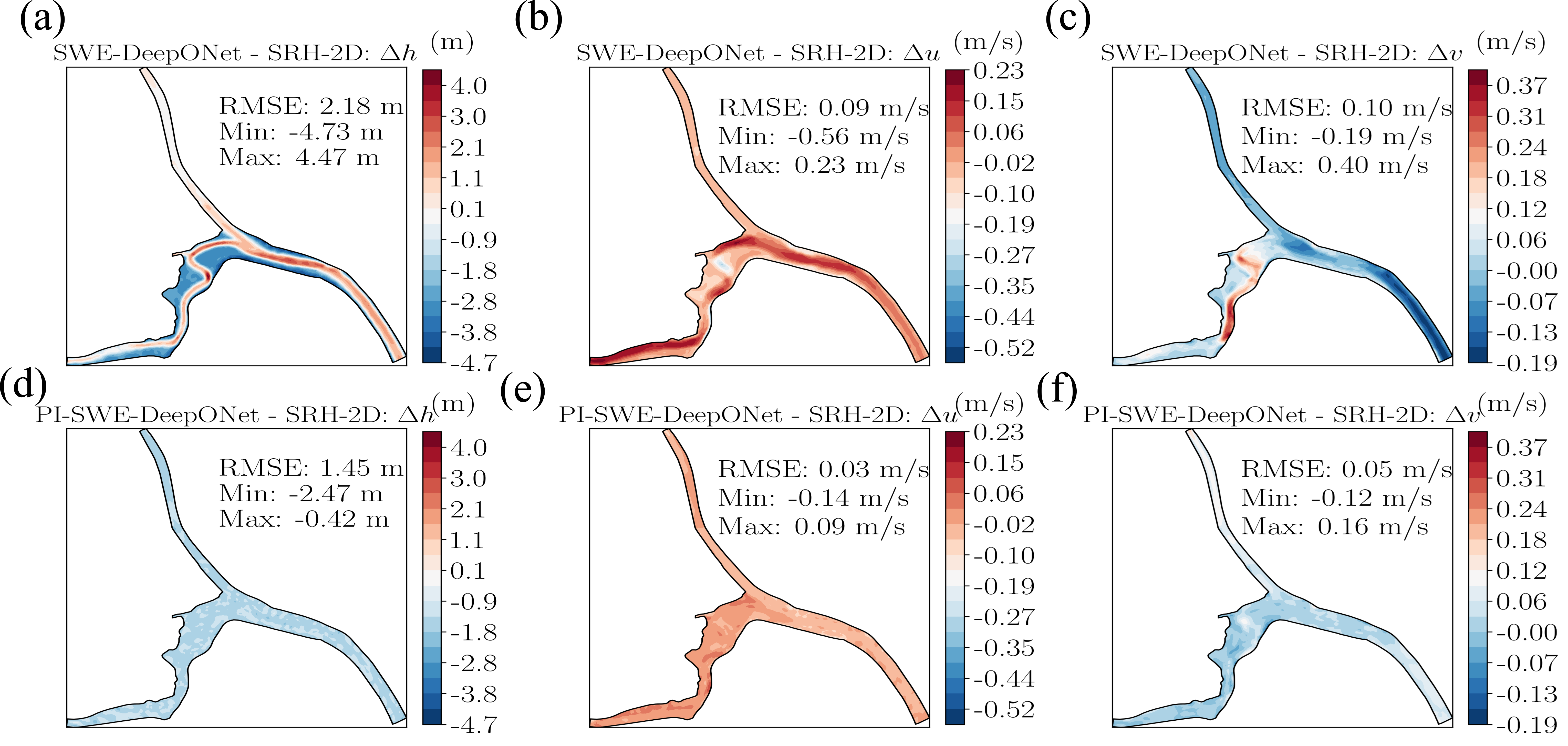}
        \caption{The differences/errors between DeepONet models and SRH-2D model for the out-of-distribution case having an average parameter distance (circled in red in Fig.~\ref{fig:comparison_w_wo_PI_difference_metrics_against_parameter_distance}): (a) $\Delta h$ between SWE-DeepONet and SRH-2D, (b) $\Delta u$ between SWE-DeepONet and SRH-2D, (c) $\Delta v$ between SWE-DeepONet and SRH-2D, (d) $\Delta h$ between PI-SWE-DeepONet and SRH-2D, (e) $\Delta u$ between PI-SWE-DeepONet and SRH-2D, (f) $\Delta v$ between PI-SWE-DeepONet and SRH-2D.}
        \label{fig:compare-application-case-diff}
    \end{figure}

\begin{figure}[htp]
    \centering
    \includegraphics[width=1\textwidth]{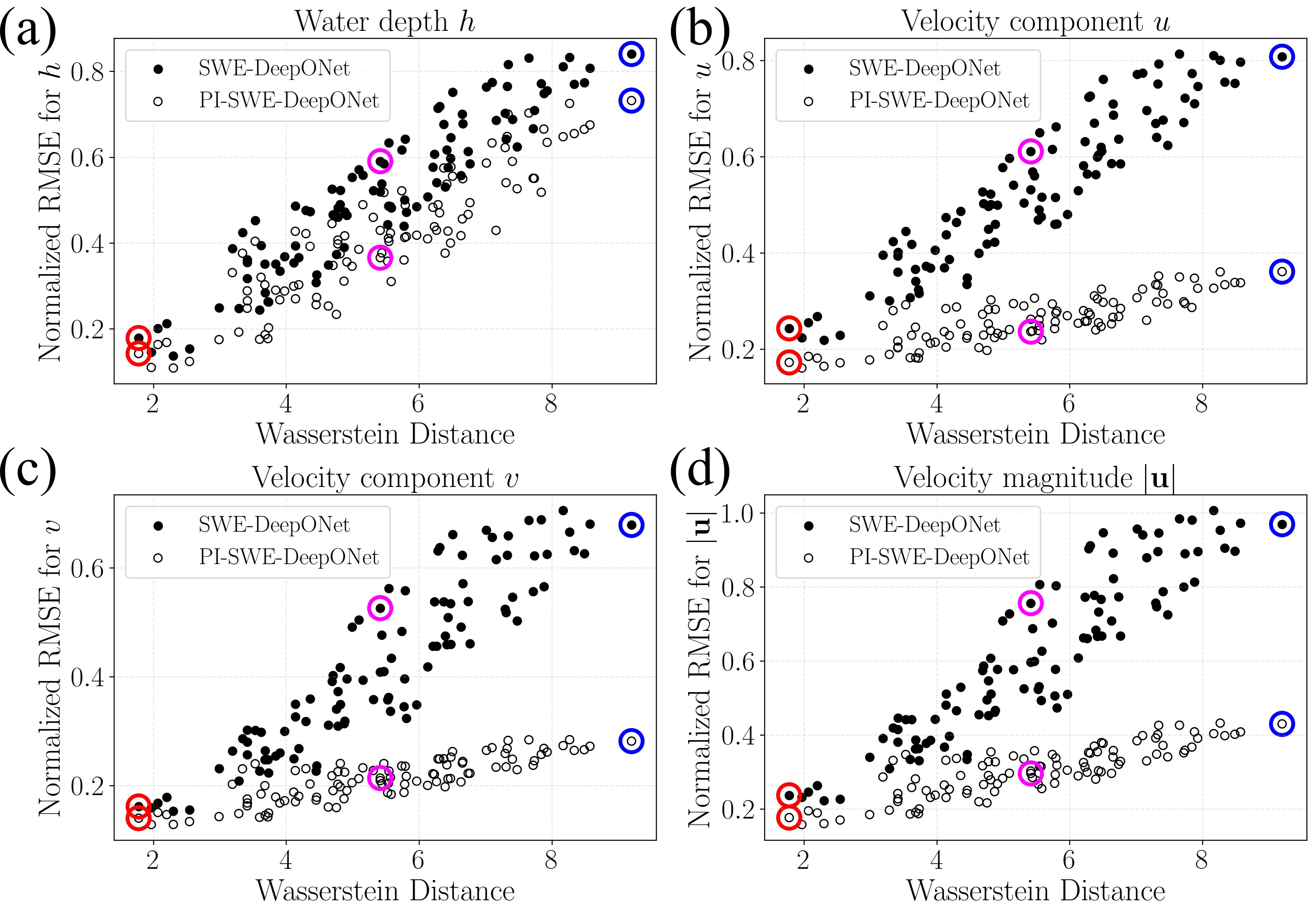}
    \caption{Performance of the SWE-DeepONet and PI-SWE-DeepONet models against SRH-2D for the out-of-distribution cases: (a) water depth $h$, (b) velocity component $u$, (c) velocity component $v$, (d) velocity magnitude $|\mathbf{u}|$. In each subfigure, the data points circled in red, magenta, and blue are those cases with minimum, average, and maximum parameter distance, respectively.}
    \label{fig:comparison_w_wo_PI_difference_metrics_against_parameter_distance}
\end{figure}

Figure~\ref{fig:compare-application-case-diff} shows a detailed comparison for a representative case with average Wasserstein distance (circled in magenta in Figures~\ref{fig:sacramento-river-inlet-discharge-pairs} and \ref{fig:comparison_w_wo_PI_difference_metrics_against_parameter_distance}). The errors of the surrogate models are computed against SRH-2D results. This case is positioned approximately in the middle of the minimum and maximum Wasserstein distance cases. It was selected as a typical example of out-of-distribution performance. The error contour maps in the figure reveal distinct differences between the two models: SWE-DeepONet has a larger error and shows more spatially extensive error patterns than PI-SWE-DeepONet. For water depth $h$, PI-SWE-DeepONet shows substantially smaller errors compared to SWE-DeepONet, with the RMSE reduced by about half. For the velocity components $u$ and $v$, PI-SWE-DeepONet exhibits smaller error magnitudes. The velocity magnitude differences follow the same trend. The consistently better performance of PI-SWE-DeepONet demonstrates the benefit of using physics constraints in maintaining physical consistency and accuracy when extrapolating beyond the training data distribution is needed.

The performance of both surrogate models with all 100 out-of-distribution cases is evaluated against SRH-2D results. The errors are plotted in Figure~\ref{fig:comparison_w_wo_PI_difference_metrics_against_parameter_distance}. Specifically, the normalized RMSEs are plotted against the Wasserstein distance for water depth $h$, velocity components $u$ and $v$, and velocity magnitude $|\mathbf{u}|$. Note that the three representative cases highlighted in Figure~\ref{fig:comparison_w_wo_PI_difference_metrics_against_parameter_distance} are highlighted: those with minimum (red), average (gren), and maximum (blue) Wasserstein distances. A key conclusion from the results is that PI-SWE-DeepONet shows a superior generalization capability in comparison with SWE-DeepONet and increased advantages are observed with increasing Wasserstein distance. This improvement is most pronounced for cases with the largest Wasserstein distances, where the physics constraints help maintain physical consistency even when the input parameters are far from the training data. 

\begin{figure}[htp]
    \centering
    \includegraphics[width=1\textwidth]{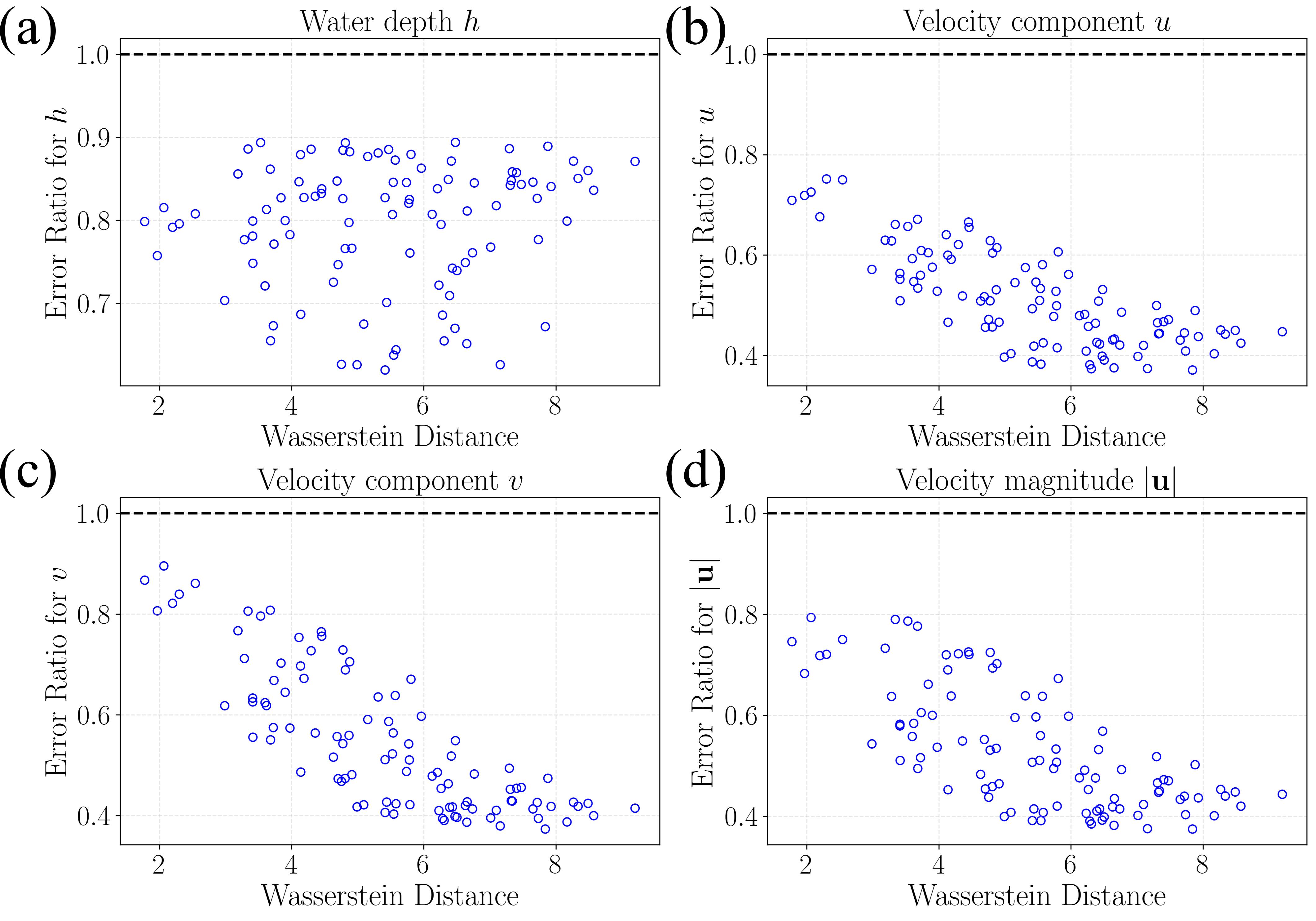}
    \caption{Error ratio between PI-SWE-DeepONet and SWE-DeepONet: (a) water depth $h$, (b) velocity component $u$, (c) velocity component $v$, (d) velocity magnitude $|\mathbf{u}|$. }
    \label{fig:comparison_w_wo_PI_improvement_ratio_against_parameter_distance}
\end{figure}

The error ratio $r$ between PI-SWE-DeepONet and SWE-DeepONet further quantifies the performance of the physics-informed model. The error-ratios plotted against the Wasserstein distance provide a clear and intuitive picture of how frequently and by how much the physics-informed model improves over the purely data-driven SWE-DeepONet model. Figure~\ref{fig:comparison_w_wo_PI_improvement_ratio_against_parameter_distance} plots the error ratio $r$ for water depth $h$, velocity components $u$ and $v$, and velocity magnitude $|\mathbf{u}|$. For all cases, the error ratio $r$ is less than 1, indicating that PI-SWE-DeepONet is more accurate than SWE-DeepONet for all out-of-distribution cases. For water depth $h$, the majority of samples fall well below unity, with error ratios typically between 0.7 and 0.9; it indicates a substantial and consistent reduction of model prediction error across the entire out-of-distribution cases. The pattern appears more pronounced and coherent for the velocity components $u$ and $v$ and velocity magnitude $|\mathbf{u}|$. The benefit due to the physics constraint tends to increase with distance ($r$ decreases with distance); it demonstrates the stabilizing effect of the physics constraints when used in the more challenging extrapolation regime. Taken together, the results show that the physics constraint provides more accurate predictions for  majority of the out-of-distribution cases and becomes increasingly beneficial as the inputs move farther away from the training distribution.

Table~\ref{tab:ood-metrics-comparison} lists the quantitative metrics described in Section~\ref{subsec:out-of-distribution-cases}. It confirms the conclusion on the advantage of using the PI-SWE-DeepONet model over SWE-DeepONet on all out-of-distribution cases. The slopes ($a_2$) for PI-SWE-DeepONet are substantially smaller than the slopes ($a_1$) for SWE-DeepONet, indicating that PI-SWE-DeepONet's error grows much more slowly with increasing Wasserstein distance. As an example, the slope for velocity component $u$ is reduced by about three quarters (0.088 → 0.024), and a reduction of about 20\% is observed for the water depth. The breakdown distances, using a threshold of normalized RMSE > 0.5, further highlight improved stability by physics constraint: for $h$, the breakdown point increases from $W_{break,1}$=4.688 to $W_{break,2}$=5.470, representing a 17\% extension of usable extrapolation range. More significant gains are obtained with velocity: the breakdown distance is increased to +$\infty$, meaning that the velocity error is always less than 0.5 for all cases. The fraction of improvement ($f_{r<1}$) reinforces these trends: PI-SWE-DeepONet outperforms SWE-DeepONet for all test samples and variables. Collectively, these quantitative metrics show that incorporating the SWE physics not only reduces overall errors but also yields substantially better robustness and reliability for out-of-distribution cases.

\begin{table}[htp]
    \centering
    \caption{Comparison of out-of-distribution performance metrics for PI-SWE-DeepONet over SWE-DeepONet.}
    \label{tab:ood-metrics-comparison}
    \renewcommand{\arraystretch}{1.3}
    \begin{tabular}{lccccc}
        \toprule
        \multirow{3}{*}{Variable} & \multicolumn{2}{c}{SWE-DeepONet} & \multicolumn{2}{c}{PI-SWE-DeepONet} & Fraction \\
        \cmidrule(lr){2-3} \cmidrule(lr){4-5}
        & Slope & Breakdown & Slope & Breakdown & of improvement \\
        & $a_1$ & Distance $W_{break,1}$ & $a_2$ & Distance $W_{break,2}$ & $f_{r<1}$ \\
        \midrule
        $h$ & 0.099 & 4.688 & 0.082 & 5.470 & 1.00 \\
        $u$ & 0.088 & 4.688 & 0.024 & +$\infty$ & 1.00 \\
        $v$ & 0.084 & 5.092 & 0.020 & +$\infty$ & 1.00 \\
        $|\mathbf{u}|$ & 0.118 & 4.140 & 0.035 & +$\infty$ & 1.00 \\
        \bottomrule
    \end{tabular}
    \renewcommand{\arraystretch}{1.0}
\end{table}

\section{Discussion}\label{section:discussion}

\subsection{On the Benefits of Using Physics-Informed Training}

An answer to the critical question of whether physics constraints are always good and when and how they should be adopted into the DeepONet training emerges from this study. This study demonstrates that the use of physical constrains is not needed for all hydraulic flows, and the choice between SWE-DeepONet and PI-SWE-DeepONet should be guided by the specific application objectives as well as the data characteristics. Physics-informed training comes with both advantages and disadvantages. The primary benefit includes enhanced generalization to out-of-distribution cases as shown in this study. Other widely known benefits include the increased training robustness when training data possesses relatively high errors or uncertainties and a reduction of the training data requirement (not studied in this work). This study shows that these benefits come at a cost: increased computing cost for the training due to the addition of physics loss computation and the increased errors for in-distribution cases. For cases with complex flow physics, such as shocks/bores and sensitive wet/dry interfaces in flat areas, PI-SWE-DeepONet may not converge well or even diverge.

Tentative guidelines may be developed based on the present study. PI-SWE-DeepONet is recommended for applications where (1) training data may contain relatively high errors or uncertainties, (2) predictions are needed for out-of-distribution scenarios, or (3) physical consistency has the priority over an accurate replication of the training data. Though not shown in this paper, there is another benefit of adding physics constraint, which is to reduce the training data requirement. It is especially useful if training data are expensive, scarce, or limited in coverage. Future work needs to explore and verify these additional benefits using the PI-SWE-DeepONet model proposed in this study.

Conversely, SWE-DeepONet is preferred when (1) applications are expected to be in-distribution, (2) training data from physics-based models is representative and highly accurate, and (3) the primary goal is to reproduce the behavior of the original physics-based model. In such cases, adding physics constraints can degrade the surrogate model performance by introducing a competing objective that prevents the model from faithfully fitting the training data derived from a physics-based solver.

\subsection{On the Competing Objectives Between Data Fidelity and Physics Constraints}

Adding PDE residuals into DeepONet loss changes the learning target fundamentally. A purely data-driven model such as SWE-DeepONet seeks to approximate the \textit{discrete} solution of SWEs produced by a physics-based hydraulic model such as SRH-2D. Once a physics constraint is added in training, the network is additionally required to satisfy the \textit{continuous} shallow water equations, which the numerical model (e.g., SRH-2D) does not satisfy exactly due to differences in the PDE formulation, discretization schemes, numerical diffusion, wet-dry treatments, approximate solution of linear equation systems, and other ad-hoc procedures to either simplify the flow or add more physics. An example ad-hoc procedure is to use the diffusive wave equations, instead of the full SWEs, at the wet/dry interface. In the present PI-SWE-DeepONet model, two inherently competing objectives are at play:
\begin{equation}
\underbrace{\text{fit the training data}}_{\parbox{4cm}{\centering approximate \textit{discrete} numerical solution of SWEs from physics-based solver}}
\qquad\text{vs.}\qquad
\underbrace{\text{minimize PDE residuals}}_{\parbox{4cm}{\centering approximate \textit{continuous} governing equations of SWEs}}
\end{equation}

Note that the training data themselves deviate from the continuous SWEs, which leads to an inconsistent optimization process during training. The result is that no network can satisfy both competing objectives. In reality, a compromise has to be made between the two.

The competing objectives are further complicated by the challenge of selecting the loss weights, which is used to balance data fidelity and physics constraints. Determining an appropriate weighting is non-trivial and subjective; the choice of different loss weighting schemes can even lead to different results. In this work, we set the weighting ratio $\lambda_{data}/\lambda_{PDEs}$ to be 2:1 based on the performance on in-distribution test cases and limited sensitivity analysis. This approach warrants further investigation because even when both losses are normalized MSE, they operate on fundamentally different scales. Specifically, the normalization schemes differ between the training data (discrete numerical solutions) and the PDE residuals (continuous governing equations), resulting in inherently different loss magnitudes that are not suitable for direct comparison.  

\section{Conclusion}\label{section:conclusion}
This study presents a physics-informed DeepONet framework for computational hydraulics modeling. Key contributions include the incorporation of the 2D shallow water equations directly into the training process, enabling the development of fast and accurate surrogate models, and demonstration of the framework using a real-world case. Preliminary guidelines were also proposed for the use of the surrogate models. The developed framework can switch on and off the SWEs physics constraint, resulting in PI-SWE-DeepONet and SWE-DeepONet models, respectively. Through application to a reach of the Sacramento River with constant-discharge flows, we demonstrate that PI-SWE-DeepONet significantly enhances model generalizability to out-of-distribution scenarios, has a slower error growth rate and larger breakdown distances in comparison with the purely data-driven SWE-DeepONet model. Despite success, however, the improved model generalization comes at a cost: slightly higher errors for in-distribution cases. This reflects the fundamental tension between two competing objectives: fitting discrete numerical solver outputs and satisfying continuous governing equations. The results show that the choice between SWE-DeepONet and PI-SWE-DeepONet models should be guided by the application requirements: PI-SWE-DeepONet is preferred when predictions are needed for out-of-distribution cases, training data contains relatively large uncertainties, or physical consistency is the priority, while SWE-DeepONet is the choice when the goal is to replicate faithfully the result of a trusted physics-based solver within the training distribution. The framework, implemented in the open-source HydroNet package, provides an automated workflow to perform a surrogate modeling: from training data generation to model training and evaluation. It makes the physics-informed operator learning accessible for practical applications. Future work should explore the potential of physics-informed training to reduce data requirements, extend the framework to unsteady flows and more real-world cases, and develop more robust loss balancing strategies to handle the competing objectives between data fidelity and physics constraints.

\section*{Open Research Section}
All code and data that support the findings of this study are available from the HydroNet GitHub repository: \url{https://github.com/psu-efd/HydroNet}.

\acknowledgments
We acknowledge the time, effort, and insights provided by the reviewers and the editor.

\bibliography{references}

\end{document}